\documentclass[twocolumn,prb,showpacs,superscriptaddress,preprintnumbers,amsmath,amssymb,floatfix]{revtex4}
\usepackage{graphicx}
\usepackage{dcolumn}
\usepackage{bm}
\usepackage{color}

\newcommand{\SRO}{SrRuO$_{3}$}
\newcommand{\STO}{SrTiO$_{3}$}

\begin{document}

\title{ Electronic and Spin States of {\SRO} Thin Films: an X-ray Magnetic Circular Dichroism Study}
\author{S.~Agrestini}
  \affiliation{Max Planck Institute for Chemical Physics of Solids,
     N\"othnitzerstr. 40, 01187 Dresden, Germany}
\author{Z.~Hu}
  \affiliation{Max Planck Institute for Chemical Physics of Solids,
     N\"othnitzerstr. 40, 01187 Dresden, Germany}
\author{C.-Y.~Kuo}
  \affiliation{Max Planck Institute for Chemical Physics of Solids,
     N\"othnitzerstr. 40, 01187 Dresden, Germany}
\author{M.~W.~Haverkort}
  \affiliation{Max Planck Institute for Chemical Physics of Solids,
     N\"othnitzerstr. 40, 01187 Dresden, Germany}
\author{K.-T.~Ko}
  \affiliation{Max Planck Institute for Chemical Physics of Solids,
     N\"othnitzerstr. 40, 01187 Dresden, Germany}
\author{N.~Hollmann}
  \affiliation{Max Planck Institute for Chemical Physics of Solids,
     N\"othnitzerstr. 40, 01187 Dresden, Germany}
\author{Q.~Liu}
  \affiliation{Max Planck Institute for Chemical Physics of Solids,
     N\"othnitzerstr. 40, 01187 Dresden, Germany}
\author{E.~Pellegrin}
  \affiliation{ALBA Synchrotron Light Source, E-08290 Cerdanyola del Vall\`{e}s, Barcelona, Spain}
\author{M.~Valvidares}
  \affiliation{ALBA Synchrotron Light Source, E-08290 Cerdanyola del Vall\`{e}s, Barcelona, Spain}
\author{J.~Herrero-Martin}
  \affiliation{ALBA Synchrotron Light Source, E-08290 Cerdanyola del Vall\`{e}s, Barcelona, Spain}
\author{P.~Gargiani}
  \affiliation{ALBA Synchrotron Light Source, E-08290 Cerdanyola del Vall\`{e}s, Barcelona, Spain}
\author{P.~Gegenwart}
  \affiliation{Experimental Physics VI, Center for Electronic Correlations and Magnetism, University of Augsburg, 86159 Augsburg, Germany}
\author{M.~Schneider}
  \affiliation{I. Physikalisches Institut, Georg-August-Universit\"{a}t G\"{o}ttingen, D-37077 G\"{o}ttingen, Germany}
\author{S.~Esser}
  \affiliation{Experimental Physics VI, Center for Electronic Correlations and Magnetism, University of Augsburg, 86159 Augsburg, Germany}
\author{A.~Tanaka}
  \affiliation{Department of Quantum Matter, ADSM, Hiroshima University, Higashi-Hiroshima 739-8530, Japan}
\author{A.~C.~Komarek}
  \affiliation{Max Planck Institute for Chemical Physics of Solids,
     N\"othnitzerstr. 40, 01187 Dresden, Germany}
\author{L.~H.~Tjeng}
  \affiliation{Max Planck Institute for Chemical Physics of Solids,
     N\"othnitzerstr. 40, 01187 Dresden, Germany}

\date{\today}
\begin{abstract}
We report a study of the local magnetism in thin films of \SRO\ grown on (111) and (001) oriented \STO\ substrates using x-ray magnetic circular dichroism spectroscopy (XMCD) at the Ru-$L_{2,3}$ edges. The application of the sum rules to the XMCD data gives an almost quenched orbital moment and a spin moment close to the value expected for the low spin state $S = 1$ . Full-multiplet cluster calculations indicate that the low spin state is quite stable and suggest that the occurrence of a transition to the high spin state $S = 2$ in strained thin films of \SRO\ is unlikely as it would be too expensive in energy.

\end{abstract}


\maketitle

Despite being investigated already for about five decades the physical properties of \SRO\ keeps fascinating the scientific community. \SRO\ is one of the few known 4$d$ transition metal oxide ferromagnets with T$_c$ as high as 160 K\cite{Callaghan66,Longo68}. Its non-integer magnetic moment has been interpreted in terms of a surprising rare example of itinerant ferromagnetism in oxides\cite{Singh96,Allen96}. More recently, the possibility of employing thin films of \SRO\ as conducting layer in epitaxial heterostructures of functional oxides has aroused a wide attention from the applied science community\cite{Koster12}.

\SRO\ is a perovskite compound with an orthorhombic GdFeO$_3$ type structure \cite{Bouchard72,Jones89}. The orthorhombic distortion arises from the zig-zag tilting, along the $c$-axis, and rotation, around the $b$-axis, of the corner-sharing RuO$_6$ octahedra. Despite this distortion the RuO$_6$ octahedra remain nearly regular \cite{Jones89,Gardner95,Bushmeleva06}. In a localized picture, the strong crystal field at the octahedral site splits the Ru 4$d$ bands of the Ru$^{4+}$ ions into  $e_g$ and $t_{2g}$ levels, leading to a low spin (LS) $t_{2g}^4$ configuration with $S = 1$. Theoretical calculations\cite{Laad01,Jeng06,Rondinelli08,Kim14} and a X-ray magnetic circular dichroism (XMCD) study\cite{Okamoto07} suggest that the orbital moment in \SRO\ should be quenched. High magnetic field measurements on a bulk single crystal give a saturated magnetization of  1.6 $\mu_B$/Ru ion\cite{Cao97}, a value similar to the ordered magnetic moment determined by neutron diffraction experiments\cite{Longo68,Bushmeleva06}.

While the technology for growing high quality \SRO\ thin films on (001) oriented \STO\ substrates, SRO/(001)STO, was developed long time ago and is well known\cite{Koster12}, the systematic growth of thin films on (111) oriented \STO\ substrates, SRO/(111)STO, is quite recent\cite{Chang09,Grutter10}. Very surprisingly the first SQUID measurements of SRO/(111)STO films have provided a saturated moment of 3.4 $\mu_B$/Ru ion \cite{Grutter10,Grutter12}, a value that is much higher than that observed in bulk \SRO\ and exceeds the atomic moment of 2 $\mu_B$/Ru ion expected for a $S = 1$ spin state. In order to explain the SQUID results it has been proposed\cite{Grutter12} that the trigonal compressive strain induced by the (111)STO substrate onto the film would stabilize the high spin state (HS) $S = 2$, which is very surprising as a HS state is unusual in $4d$ oxides. Even more intriguing, an unquenched orbital moment of about 0.32 $\mu_B$ has been reported for these strained films on the basis of XMCD measurements at the Ru $M_{2,3}$ edges\cite{Grutter12}. However, theoretical studies, which investigated the effect of substrate-induced compressive strain on the physical properties of \SRO\, could not find evidence in support of the alleged stabilization of a HS state or even suggested the reduction of the magnetic moment from bulk values\cite{Kim14,Zayak08}. Further, a very recent study\cite{Lee14} on SRO/(111)STO has reported magnetization values suggesting a LS state, in contradiction with the results published earlier\cite{Grutter10,Grutter12}. Understanding the stability of the magnetic ground state of \SRO\ is obviously a very important aspect for controlling the magnetic properties of heterostructures involving \SRO\ as conducting layer.

\begin{figure}[ht]\centering
\includegraphics[width=6 cm]{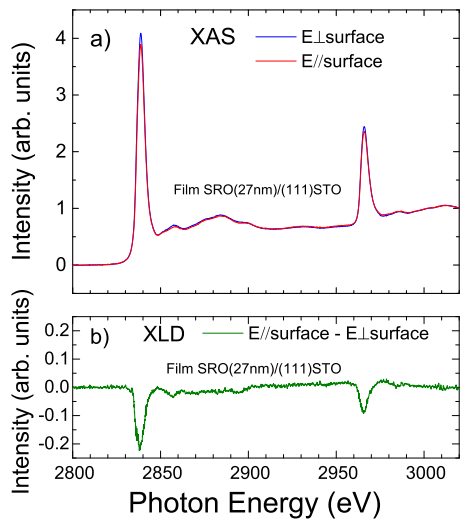}
\caption{Ru-$L_{2,3}$ XAS spectra of a 27 nm SRO/(111)STO film for linearly polarized light coming in with the electric field vector $\textbf{E}$ normal [dark (blue) lines] and parallel [light (red) lines] to the film surface.}\vspace{-0.2cm}
\end{figure}

In this work we address two questions: 1) whether compressive strain can induce a spin state transition in SRO/(111)STO and 2) whether the orbital moment is quenched. To this end, we have performed an investigation of the XMCD signal at high magnetic field at the Ru $L_{2,3}$ edges of \SRO\ films under different compressive strains (trigonal strain for the case of (111)STO substrate and tetragonal strain for the case of (001)STO substrate) compared with the case of a \SRO\ single crystal. 
XMCD is a well-established technique to study local magnetic properties. The XMCD signal can be analyzed by means of sum rules\cite{thole92a,Carra93}, allowing for a direct experimental determination of the desired quantum numbers $L_z$ and $S_z$. The energy separation between Ru $L_{3}$ and $L_{2}$ edges of about 150 eV is much larger than the multiplet effects (a few eV), and therefore the spectra are very suitable also for spin sum rule analysis\cite{Carra93}. In addition, the signal-to-background ratio at the $L_{2,3}$ edges is higher than at the $M_{2,3}$ edges. We would like to stress that obtaining a reasonable degree of circular polarized light at the photon energies of the $L$ edges of $4d$ elements is challenging\cite{Tomaz98} and only thanks to the development of the new BOREAS beamline this XMCD investigation of the Ru $L_{2,3}$ edges has been possible. In addition, a comparison of the line shape to full-multiplet theory can be made to unravel details of the wave functions forming the ground state.

Single crystalline thin films of \SRO\ were grown on SrTiO$_3$ substrates with different orientations by metalorganic aerosol deposition. Thin \SRO\ films grown on (001) and (111) oriented substrates were determined by X-ray diffraction (XRD) to have (100)c and (111) orientation, respectively. (In this report, we use pseudocubic notation for SRO films. (110)orthorhombic and (101)orthorhombic is equivalent to (100)c and (111) in the pseudocubic notation).  The XRD results show that the films grown on (111) oriented substrates exhibit an elongation of the out-of-plane lattice constant (3.946(1) and 3.950(1) {\AA} for the 80 and 27 nm thick films, respectively) compared to bulk \SRO\ ($\simeq3.93$ {\AA}\cite{Grutter12}). This systematic evolution of the out-of-plane constant with film thickness (the thinner the film, the larger the out-of-plane constant) is an effect of the strain: under compressive in-plane strain the in-plane lattice constant shrinks, while the out-of-plane lattice constant becomes elongated, in order to roughly preserve the unit cell volume\cite{Lee14}. The thickness of the \SRO\ films was determined by small-angle X-ray scattering.
Details of their preparation and structure characterization are reported in Ref.~24. Large single crystals of \SRO\ were grown by floating zone technique. The purity and quality of the crystal were checked by x-ray diffraction. Susceptibility measurements using a MPMS squid magnetometer show a bulk ferromagnetic transition at $T_c$ = 160~K for the single crystal, and between 154~K and 147~K for the films depending on the film thickness.
The x-ray linear dichroism (XLD) and x-ray magnetic circular dichroism (XMCD) experiments at the Ru-$L_{2,3}$ edges (2800-3000 eV) were performed at the BL29 Boreas beamline at the ALBA synchrotron radiation facility in Barcelona. The energy resolution was 1.4~eV and the degree of circular polarization delivered by the Apple II-type elliptical undulator was adjusted to 70\% as balanced trade off between degree of polarization and photon flux required when working at high photon energies and high undulator harmonics. The degree of linear polarization for XLD is close to 100\%. The XMCD signal was measured using a magnetic field of 6~Tesla with the sample at a temperature of 50~K. The spectra were recorded using the total electron yield method (by measuring the sample drain current) in a chamber with a vacuum base pressure of 2x10$^{-10}$~mbar. The single crystalline sample was cleaved in situ to obtain a clean sample surface normal to the (110) direction. The XAS spectra were collected in both $B$ = 6~T and -6~T applied fields and in groups of four or quartet ($\sigma^+\sigma^-\sigma^-\sigma^+$ or $\sigma^-\sigma^+\sigma^+\sigma^-$, where $\sigma^+$ and $\sigma^-$ indicate photon spin parallel or antiparallel to the applied field, respectively) in order to minimize the effect of any time dependence in the X-ray beam on the measured spectra.

\begin{figure}[ht]\centering
\includegraphics[width=6.5 cm]{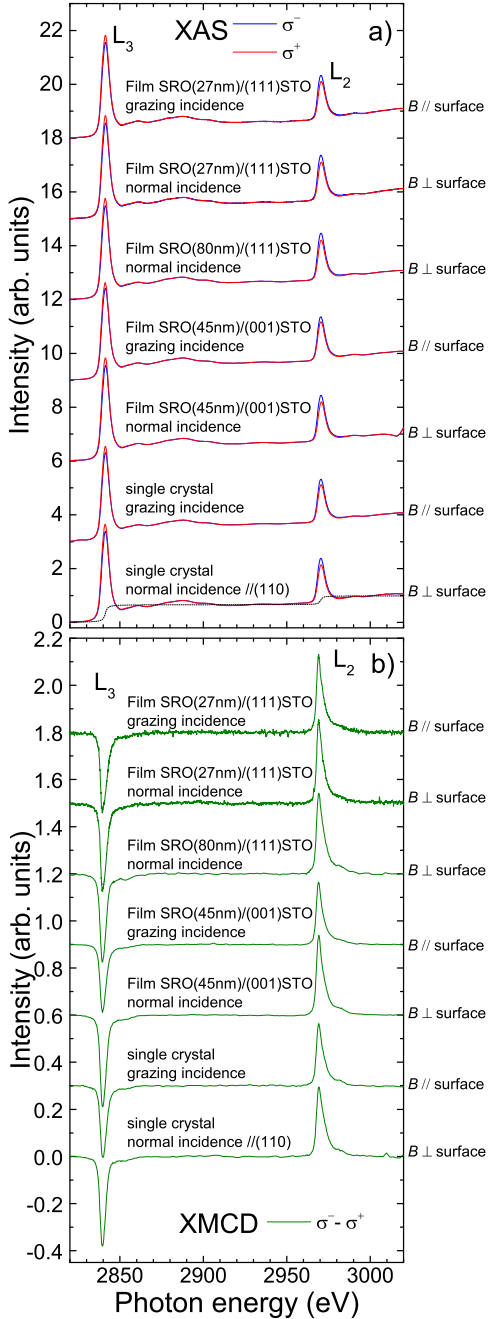}
\caption{Ru-$L_{2,3}$ XAS spectra a) and XMCD spectra b) of \SRO\ films and bulk single crystal measured at $T$ = 50~K and H = 6~T. The spectra are vertically shifted for clarity. The spectra taken at normal (90$^o$) and grazing (20$^o$) incidence show a small anisotropy. The dashed black curve represents the edge jump.}\vspace{-0.2cm}
\end{figure}

In Fig. 1 we report the Ru-$L_{2,3}$ XAS measured on a 27~nm SRO/(111)STO film for linearly polarized light coming in with the electric field vector $\textbf{E}$ normal [dark (blue) lines] and parallel [light (red) lines] to the film surface. The Ru $2p$ core-hole spin-orbit coupling splits the spectrum roughly in two parts, namely the $L_{3}$ (at $h\nu \approx$~2840~eV) and $L_{2}$ (at $h\nu \approx$~2970~eV) white lines regions. A clear linear dichroism (XLD) can be observed, which is an indication that the film is under in-plane compressive strain. In fact, in-plane compressive strain leads to a trigonal elongation of the RuO$_6$ octahedron along the (111) axis. As consequence, the $t_{2g}$ orbitals are split in $a_{1g }$ and $e_g^{\pi}$ orbitals, with the $a_{1g }$ orbital lying higher in energy and, hence, having more holes. The experimentally observed larger spectral weight for $\textbf{E}$ normal to the film surface is a result of the uneven hole distribution among the $t_{2g}$ orbitals induced by the strain.

The top panel of Fig. 2 shows the Ru-$L_{2,3}$ XAS measured on SRO/(111)STO and SRO/(001)STO films and, for comparison, on a \SRO\ single crystal. The XAS spectra were taken using circular polarized light with the photon spin parallel ($\sigma^+$, red curves) and antiparallel ($\sigma^-$, blue curves) aligned to the magnetic field. The difference spectrum ($\sigma^--\sigma^+$), i.e., the XMCD spectrum, is reported in the bottom panel of Fig. 2. The spectra were collected with the beam in grazing ($B//$surface) and in normal ($B\bot$surface) incidence, see Fig. 3 for experimental geometry. The XMCD signal is larger for $B\bot$surface than for $B//$surface by about 30\% for the SRO/(001)STO film, and by about 5\% for the 27 nm SRO/(111)STO film. The anisotropy of the XMCD signal agrees with the picture of an out-of-plane easy axis for \SRO\ films grown on STO as reported in literature \cite{Grutter10,Lee14}. The reduced magnetic anisotropy shown by our XMCD measurements in the case of SRO/(111)STO films with respect to SRO/(001)STO film is in fair agreement with previous SQUID measurements\cite{Lee14}. Both XAS and XMCD spectra measured on the SRO/(111)STO and SRO/(001)STO films appear fairly identical to those measured on the bulk single crystal, without clear evidence of changes in the spectral lineshape and in the size of the XMCD signal that otherwise could suggest a different spin state.

\begin{figure}[t]\centering
\includegraphics[width=6 cm]{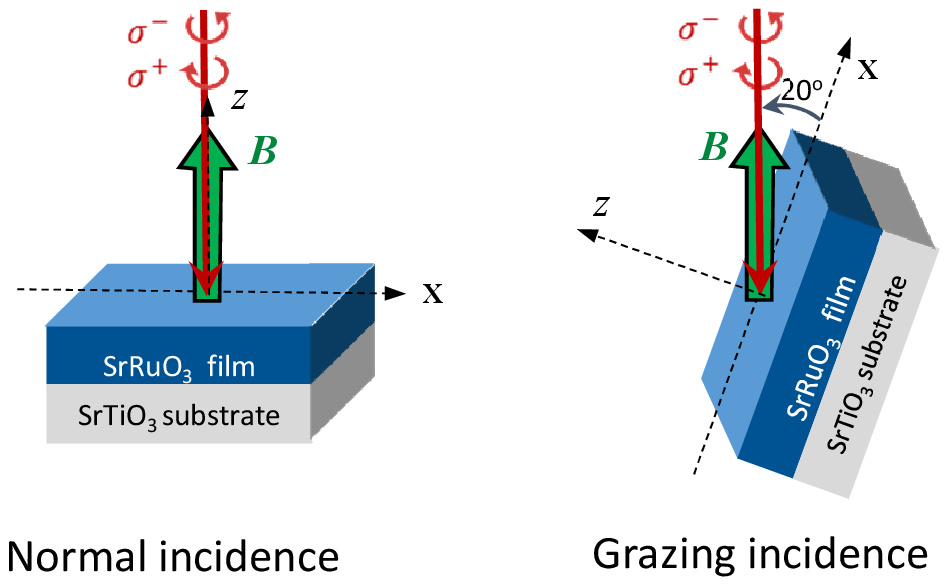}
\caption{Setup of the XMCD experiments: the magnetic field $B$ is applied parallel to the Poynting vector of the circularly polarized photons and forms an angle of 90$^o$ (20$^o$) in normal (grazing) incidence with the sample surface. }\vspace{-0.2cm}
\end{figure}

\begin{table}[ht]
	\centering
		\begin{tabular} {c c c c c c c c}\hline\hline
		& Sample& incidence & $L_z/2S_z$ &  $L_z$ & $2S_z$ \\	\hline
		& Crystal & $B\bot$surface\ & 0.01 & 0.01& 1.9 \\
		& Crystal & $B//$surface & 0.01 & 0.02& 1.7 \\
		& SRO(45nm)/(001)STO & $B\bot$surface & 0.01 & 0.01& 1.9 \\
		& SRO(45nm)/(001)STO  & $B//$surface & 0.01 & 0.02& 1.5 \\
		& SRO(80nm)/(111)STO  & $B\bot$surface & 0.00 & 0.00& 2.0 \\
		& SRO(27nm)/(111)STO  & $B\bot$surface & 0.01 & 0.03& 1.9 \\
		& SRO(27nm)/(111)STO  & $B//$surface & 0.01 & 0.02& 1.8 \\
			\hline\hline
		\end{tabular}
\caption{L/2S ratio, orbital and spin moment as estimated using sum rules. The values were divided by a factor 0.7 to take into account that the beam was only 70\% circular polarized.}
\label{Data35K}
\end{table}

The material metallicity and life time broadening ($\thicksim$2 eV) may limit the information that can be obtained from the lineshape about the Ru ground state. However, it is possible to use the sum rules for XMCD developed by Thole and Carra et al. \cite{thole92a,Carra93}  to extract from our XMCD data the orbital ($L_z$) and spin (2$S_z$) moments:
\begin{equation}\label{eq:lz}
L_z=\frac{4}{3}\cdot \frac{\int_{L_{2,3}}(\sigma^+-\sigma^-)dE}{\int_{L_{2,3}}(\sigma^++\sigma^-)dE} \cdot N_h,
\end{equation}

\begin{equation}\label{eq:ratio}
2S_z+7T_z=2\cdot \frac{\int_{L_{3}}(\sigma^+-\sigma^-)dE-2\int_{L_{2}}(\sigma^+-\sigma^-)dE}{\int_{L_{2,3}}(\sigma^++\sigma^-)dE} \cdot N_h,
\end{equation}

For ions in octahedral symmetry the magnetic dipole moment $T_z$ is a small number and can be neglected compared to $S_z$ \cite{Teramura96}. The number of holes in the $4d$ shell was estimated to be about $N_h=5.2$ by our cluster calculations, in agreement with previous estimates\cite{Guedes12}, reflecting the highly mixed $p-d$ covalency of the ground state in \SRO. In estimating the XAS intensities, the edge jump background, described as arctan function, has been subtracted from the XAS spectra (dashed curve in Fig. 2). The results of the application of the sum rules are reported in table 1. The orbital moment $L_z$ is found to be almost quenched for all samples, including the SRO/(111)STO. The spin contribution to the magnetic moment in the 80 and 27 nm SRO/(111)STO films in normal incidence is found to be close to the value expected for a $S=1$ spin state and very similar to that found for the bulk single crystal\cite{spin}. These results are in clear contradiction with the much larger saturated moment values reported earlier\cite{Grutter12} from SQUID measurements on SRO/(111)STO films.

\begin{figure}[t]\centering
\includegraphics[width=8 cm]{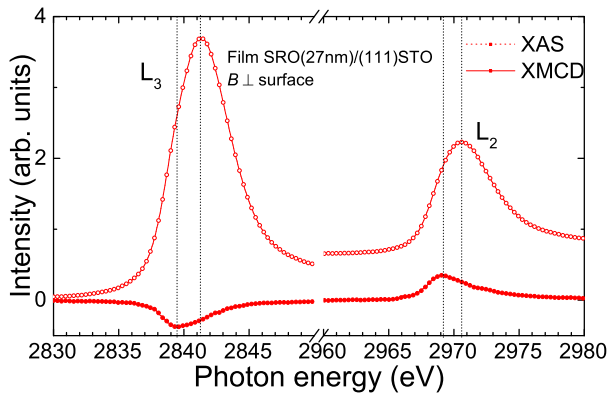}
\caption{Zoom of Ru-$L_{2,3}$ isotropic XAS and XMCD spectra of a 27 nm thin film of \SRO\ grown on (111) oriented \STO\ substrate. The vertical dashed lines corresponds to the energy position of the maxima in the XAS and XMCD spectra.} \vspace{-0.2cm}
\end{figure}

As mentioned before, the spectra seem to be rather featureless, but a closer look reveals that for both $L_{3}$ and $L_{2}$ edges the maximum in intensity of the XAS spectrum lies ~1.5 eV higher in energy position than that of the XMCD spectrum (see Fig. 4). We see exactly the same difference in energy position in the spectra of all samples.
A similar energy shift of the XMCD peak with respect to the XAS peak was previously observed for the Ru-$M_{2,3}$ edges\cite{Okamoto07} and can be understood considering that only the $t_{2g}$ orbitals contribute to the XMCD signal, while both $t_{2g}$ and $e_g$ orbitals contribute to the XAS spectrum with the XAS maximum corresponding to the signal from the unoccupied $e_g$ levels. Therefore, this energy position difference provides a very important information as it reflects the crystal field splitting $10Dq$ between the $t_{2g}$ and $e_g$ orbitals. In order to determine quantitatively $10Dq$ we have performed simulations of the XAS and XMCD spectra using the well-proven full-multiplet configuration-interaction approach\cite{degroot94, thole97}. It  accounts for the intra-atomic $4d-4d$ and $2p-4d$ Coulomb interactions, the atomic $2p$ and $4d$ spin-orbit couplings, the oxygen $2p-4d$ hybridization, and local crystal field parameters. In the simulations we considered a RuO$_6$ cluster with a cubic symmetry as the octahedra in bulk \SRO\ are fairly regular\cite{Jones89,Gardner95,Bushmeleva06}. The calculations were performed using the XTLS 8.3 code\cite{Tanaka94} with the parameters given in Ref. 31. We applied to the spectra additional Gaussian (1.4 eV) and Lorentzian (2 eV) broadening in order to take into account experimental resolution and lifetime effects, respectively.

\begin{figure}[t]\centering
\includegraphics[width=6 cm]{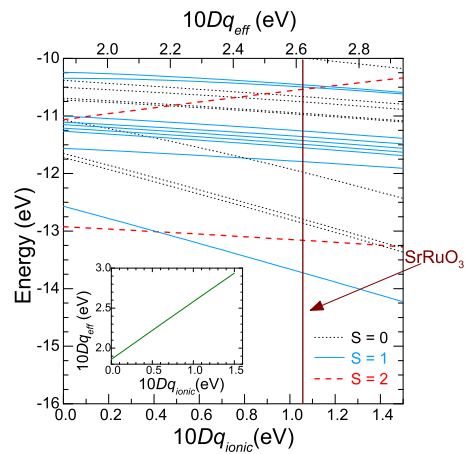}
\caption{Energy level diagram of the Ru$^{4+}$ ion as a function of the ionic crystal field $10Dq$ in a cubic local symmetry. Black dotted, blue solid and red dashed lines correspond to levels with $S =$ 0, 1 and 2 spin state, respectively. The inset shows the evolution of the effective crystal field versus ionic crystal field. The vertical magenta solid line indicates the $10Dq_{eff}$ of \SRO\ as obtained by the simulation of the XAS and XMCD spectra.} \vspace{-0.2cm}
\end{figure}

\begin{figure}[t]\centering
\includegraphics[width=9 cm]{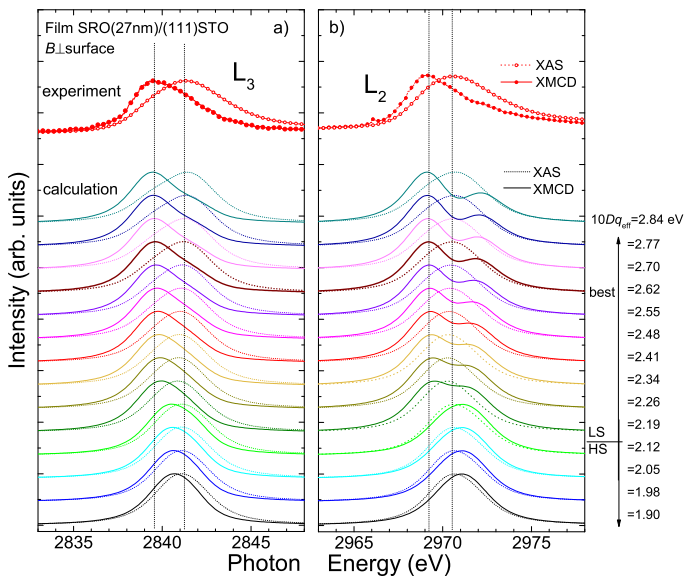}
\caption{Experimental (top) and simulated (below) XMCD and XAS spectra for different $10Dq_{eff}$ values at the $L_3$ (panel a) and $L_2$ (panel b) edges. For a better comparison the spectra were normalized to the peak height and the Ru $L_3$ XMCD spectra was reversed. The vertical dotted lines correspond to the energy position of the maxima in the XAS and XMCD spectra.} \vspace{-0.2cm}
\end{figure}

In Fig.~5 we report the energy level diagram of the Ru$^{4+}$ ion as a function of the ionic crystal electric field 10$Dq_{ionic}$  in a cubic local symmetry. The energy difference between the two $S = 2$ levels with orbital occupation $t_{2g}^3e_g^1$ and $t_{2g}^2e_g^2$ (bottom and top red dashed line in Fig.~5, respectively) can be taken as a measure of the effective crystal electric field $10Dq_{eff}$, i.e. the splitting between $t_{2g}$ and $e_g$ levels including the effect of the hybridization with the oxygens. The diagram shows that for $10Dq_{eff} >$~2.15~eV ($10Dq_{ionic} >$~0.41~eV) the level with configuration $t_{2g}^4e_g^0$ is the lowest energy level (bottom solid blue line) and the ground state of Ru$^{4+}$ ion has $S = 1$ spin state. As the crystal field is reduced across the critical value of $10Dq_{eff} \leq$~2.15~eV ($10Dq_{ionic} \leq$~0.41~eV) the $t_{2g}^3e_g^1$ level (red dashed line) becomes the lowest energy level and the HS state is stabilized. The non-magnetic $S = 0$ state (black dotted lines) lies always much higher in energy and never becomes the ground state for any value of the cubic crystal field.

In Fig. 6 we show the comparison of the simulated XAS and XMCD spectra at the $L_{2,3}$ edges with the experimental spectra measured on SRO/(111)STO film. The simulated XAS and XMCD spectra were calculated for different values of the ionic and effective crystal field splitting. For the sake of clarity, the spectra were normalized to the height of the peak and the XMCD signal at the $L_3$ edge was reversed. The calculated spectra show that the peak position depends on the value of crystal field splitting. The experimental energy separation between the maxima of the XMCD and XAS spectra can be correctly simulated for $10Dq_{eff} =$~2.62~eV and the lineshape of the calculated spectra is fairly similar to that of the experimental spectra. For such a value of $10Dq_{eff}$ the Ru$^{4+}$ ions are in a LS $S = 1$ ground state. The HS spin state becomes stable only for smaller crystal field splitting, $10Dq_{eff} <$~2.15~eV. In the hypothesis of a HS spin state as a ground state the simulated XMCD spectrum looks very different from the experimental one: 1) at the $L_3$ and $L_2$ edge the XMCD lineshape is not anymore asymmetric; 2) at the $L_2$ edge the XMCD maximum occurs at higher photon energy than the XAS maximum, which is opposite to what has been experimentally observed. As it can be seen in the energy level diagram reported in Fig. 5 \SRO\ is located very far from the stability region for the HS state.

To summarize, we have used XMCD spectroscopy to investigate the local magnetism in thin films of \SRO\ grown on (111) and (001) oriented \STO\ substrates. We have found that the orbital moment is almost quenched and the spin is close to the value expected for a $S = 1$ spin state. From a comparison of the experimental with simulated spectra we could determine the effective crystal field. The hypothesis of a compressive strain-induced spin state transition, as proposed in literature on the basis of SQUID measurements, can be ruled out as the stabilization of the high spin state with $S = 2$ would be too costly in energy.

The XMCD experiments were performed at the BOREAS beamline of ALBA Synchrotron with the collaboration of ALBA staff. The research leading to these results has received funding from the European Community's Seventh Framework Programme (FP7/2007-2013) under grant agreement n.$^o$312284. Q. Liu received financial support from the European Union through the ITN Soprano Network (Grant No. PITN-GA-2008-214040). K.-T. Ko acknowledges support from the Max Planck-POSTECH Center for Complex Phase Materials (No. KR2011-0031558).


\begin{thebibliography}{40}
\bibitem{calc_par} $U_{dd}$ = 3.0 eV, $U_{pd}$ = 4.0 eV, $pd\sigma$ = 2.19 eV, $pd\pi$ = -1.03 eV and $\Delta$ = 1.0 eV. Slater integrals were reduced to 70~\% of the Hartree-Fock values. The Ru 4$d$ spin-orbit parameter was set to zero to avoid the $J = 0$ (singlet) state to form the ground state, and instead, to allow the $S = 1$ or $S = 2$ states to become stable.
\end{thebibliography}
\end{document}